\documentstyle[prl,aps, multicol,epsf]{revtex}

\begin{document}
\draft

\title{Evaporation of the pancake-vortex lattice in
weakly-coupled layered superconductors}

\author{
  M.\ J.\ W.\ Dodgson,$^{a,*\,}$ 
  A.\ E.\ Koshelev,$^{b\,}$ 
  V.\ B.\ Geshkenbein,$^{a\,}$  
  and G. Blatter$^{a\,}$ 
}

\address{$^{a\,}$ Theoretische Physik, ETH-H\"onggerberg, CH-8093
Z\"urich, Switzerland}

\address{$^{b\,}$ Materials Science Division, Argonne National
  Laboratory, 9700 S.\ Cass Ave, Illinois 60439, USA}

\date{\today} \maketitle
\begin{abstract}

We calculate the melting line of the pancake-vortex system in a
layered superconductor, interpolating between two-dimensional (2D)
melting at high fields and the zero-field limit of single-stack
evaporation. Long-range interactions between pancake vortices in
different layers permit a mean-field approach, the ``substrate
model'', where each 2D crystal fluctuates in a substrate potential due
to the vortices in other layers.  We find the thermal stability limit
of the 3D solid, and compare the free energy to a 2D liquid to
determine the first-order melting transition and its jump in entropy.

\end{abstract}

\pacs{PACS numbers:
74.60.Ec, 
74.60.Ge, 
63.70.+h, 
64.70.Dv  
}

\begin{multicols}{2}
\narrowtext


The pancake-vortex lattice in layered superconductors defines a
tunable soft matter system with astonishing properties \cite{review}.
Among them, the thermodynamic phase transition of vortex-lattice
melting and its first-order character is now experimentally well
established \cite{Zeldov,Schilling}, but questions remain as to which
correlations are lost at the transition \cite{Fuchs}. Theoretically,
the position of the melting line can be estimated with a Lindemann
criterion \cite{Houghton,BlatterGLN}, but a more detailed description
of melting is required to determine the characteristics of the
transition. The challenge in defining a theoretical scheme describing
vortex-lattice melting follows from the complexity of the vortex
system in real superconductors combined with the general lack of exact
theories of melting.

\vspace{-0.10cm} In a moderately anisotropic material, such as
YBa$_2$Cu$_3$O$_7$, the vortex crystal melts to a line liquid and
numerical simulations have treated this in
detail\cite{LargeLamSims}. In Bi$_2$Sr$_2$CaCu$_2$O$_8$ (BSCCO),
however, the coupling between layers is so weak that the layered
structure (with spacing $d$) plays a crucial role, and the vortex
matter acts as a collection of interacting two-dimensional (2D)
vortices, or pancake vortices.  Rather than using numerical
simulations \cite{Reefman,Ryu}, we describe here a novel analytic
treatment to track the melting line through the $B$-$T$ phase diagram
in the extreme anisotropic limit of zero Josephson coupling between
layers.  In this limit the 3D pancake-vortex lattice (PVL) remains
stable at low temperatures due to an attractive electromagnetic
interaction between pancake vortices in different layers, with range
$\lambda\gg d$ ($\lambda$ is the in-plane penetration depth). Changing
the magnetic field tunes the relative importance of this attractive
interlayer interaction and the long-range repulsion between vortices
in the same layer. At high fields $B\gg B_\lambda=\Phi_0/\lambda^2$,
the in-plane interactions dominate and the 3D lattice melts to
independent 2D liquids (a pancake-vortex gas) close to the 2D melting
temperature $T_m^{\rm\scriptscriptstyle 2D}\approx \varepsilon_0 d/70$
\cite{Caillol} [where $\varepsilon_0=(\Phi_0/4\pi\lambda)^2$]. At
lower fields the interlayer attraction stabilizes the lattice and
increases the melting temperature. In the low-field limit of
weakly-coupled 1D stacks, the crystal melts below the evaporation
transition of an isolated stack of pancake vortices \cite{Evaporation}
located at the Berezinskii-Kosterlitz-Thouless (BKT) vortex unbinding
transition\cite{KT} of an isolated layer at $T_{\rm\scriptscriptstyle
BKT}=\varepsilon_0d/2\sim 35 T_m^{\rm\scriptscriptstyle 2D}$.  The
field regime $B\sim B_\lambda$ where the melting line interpolates
between the above limits then spans a factor $\sim 35$ in ``reduced''
temperature $T/\varepsilon_0d$ (in real superconductors
$\varepsilon_0$ vanishes as $T_c$ is approached, and the real
temperature ratio $T_{\rm\scriptscriptstyle
BKT}/T_m^{\rm\scriptscriptstyle 2D}$ will be smaller).  Before
reaching zero field, the melting line is cut by a competing low-field
reentrant transition to a dilute liquid of stacks with
exponentially-weak interactions\cite{review}. A Lindemann
analysis\cite{BlatterGLN} tells us that at $T\sim\varepsilon_0 d/2$
reentrant melting occurs at a field below $10^{-2}B_\lambda$.

\vspace{-0.05cm} Ignoring reentrance, we have a 3D melting line that
interpolates between a 1D stack evaporation (exactly described by 2D
BKT theory) and a 2D melting transition usually described by a
BKT-type mechanism of dislocation pair unbinding. Both limits are well
described by a self-consistent approximation and we here generalise
this to all magnetic fields. Our self-consistent method relies on the
long range of the inter-layer attractions; each pancake vortex feels
the attractive force of pancake vortices in $\sim\lambda /d$ other
layers. Therefore the fluctuations in pancake-vortex positions may be
averaged, leading to an accurate ``mean-field'' approach where the 2D
lattice in one layer sits in a substrate potential due to the
attraction of the vortex stacks in all other layers. With this {\em
substrate model} we calculate the fluctuations of individual pancake
vortices, which in turn smears the substrate potential, and we solve
self-consistently. The upper bound in temperature to a self-consistent
solution leads to an instability line which we calculate in this
paper. We then determine the melting line by comparing the free energy
of the 3D PVL to the free energy of a collection of 2D liquids, using
numerical results for the 2D system\cite{Caillol}.

\vspace{-0.05cm} The evaporation at $T_{\rm\scriptscriptstyle BKT}$ of
a single stack of pancake vortices occurs because each element is only
logarithmically bound to the stack: A pancake vortex in a layered
superconductor generates supercurrents within each layer, resulting in
a pairwise interaction energy \cite{Pancakes},
\begin{equation}
  \label{eq:interaction}
  V_n({\bf R})=
  \frac{\Phi_0^2d^2}{4\pi\lambda^2}\!\int\!\frac{d^2Kdk_z}{(2\pi)^3} 
  \frac{(K^2+k_z^2)e^{i(k_znd+{\bf K}\cdot{\bf R})}}{K^2
(\lambda^{-2}+K^2+k_z^2)},
\end{equation}
where $n$ is the number of layers separating the two vortices, and
${\bf R}$ is the in-plane distance.  The form of this interaction
for different limits is well documented\cite{review}: the
in-plane repulsion is $V_0(R)=-2\varepsilon_0 d
\,\ln(R/L)$ (where L is the system-size cut off) and the
out-of-plane attraction has the large $R$ limit, $V_{n\ne0}(R)
=\varepsilon_0 d (d/\lambda) e^{-nd/\lambda}\ln(R/L)$. 
This implies that the energy to pull a single pancake vortex (of 
core-size $\xi$)
from a straight stack is $2\varepsilon_0 d \,\ln(R/\xi)$ 
when $R\gg\lambda$ and the entropy will unbind the pancake 
vortices above $T_{\rm\scriptscriptstyle BKT}$
\cite{Evaporation}.

This stack evaporation is easily reproduced within a self-consistent
substrate model \cite{Dodgson_LT22}. Here, each pancake vortex is
subject to a quadratic potential, but with a strength chosen to match
the thermal average ($\langle \dots \rangle$) of the curvature in the
real potential\cite{Dodgson_preprint}, $V_{\rm s}({\bf
u}_0)=\frac{1}{2}\alpha_{\rm s}{u_0}^2$, where ${\bf u}_n$ is the
$n$-th pancake vortex displacement, and
\begin{equation}
\alpha_{\rm s} = \sum_{n\ne 0} \left\langle 
\frac{\partial^2 V_n({\bf u}_n-{\bf u}_0)}{\partial u_0^x 
\partial u_0^x} \right\rangle.
\end{equation}
We ignore correlations in the pancake vortex fluctuations and use the
identity for Gaussian fluctuations that $\left\langle \exp[-i{\bf
K}\cdot({\bf u}_n-{\bf u}_0)]\right\rangle =\exp{(-K^2\langle
u^2\rangle/2)}$, to give
\begin{equation}
\alpha_{\rm s} = -\sum_{n\ne 0} \int\frac{d^2K}{(2\pi)^2}
{K_x}^2  V_n(K) e^{-\frac{K^2\langle u^2\rangle}{2}},
\label{eq:phiofu2}
\end{equation}
where $V_n(K)=\int d^2R e^{-i{\bf K}\cdot{\bf R}}V_n(R)$. The
equipartition theorem for a harmonic potential, $\langle
u^2\rangle=2T/\alpha_{\rm s}$, allows us to solve
Eq.~(\ref{eq:phiofu2}) self-consistently: for large displacements
$\langle u^2\rangle\gg\lambda^2$ the limiting form is $\alpha_{\rm s}
=\varepsilon_0d/(\langle u^2\rangle+2\lambda^2)$, which has the
solution $\langle u^2\rangle=2\lambda^2/[1-(2T/\varepsilon_0d)]$,
diverging at the evaporation temperature $T_{\rm\scriptscriptstyle
BKT} = \varepsilon_0d/2$.

We now extend this self-consistent analysis to the full 3D system at
finite fields. We consider the full 2D fluctuations of the crystal in
each layer, sitting on a substrate due to the stacked vortex crystals
in the other layers. Before deriving this in detail, we give a
quick-and-dirty derivation of evaporation at small fields. Close
enough to $T_{\rm\scriptscriptstyle BKT}$ the instability occurs when
$\lambda^2\ll \langle u^2\rangle\ll a_0^2$, for a vortex density
$n_v=\Phi_0/B=2/\sqrt{3}a_0^2$. In this limit the substrate potential
picks up a negative background contribution (see below),
$\alpha_s\approx \varepsilon_0d\left[-2\pi n_v+1/(\langle
u^2\rangle+2\lambda^2)\right]$. Inserting this to the equipartition
result gives the quadratic equation in $\langle u^2\rangle$, $\langle
u^2\rangle=(2T/\varepsilon_0 d) [2\lambda^2 + \langle u^2\rangle +
2\pi n_v \langle u^2\rangle^2]$, which only has solutions below a
temperature given by $\left(1-T_{\scriptscriptstyle \rm
BKT}/T\right)^2 -16\pi n_v\lambda^2=0$, and the instability line
approaches the zero-field transition in the form
\begin{equation}
B_u\sim B_\lambda \left(1-T/T_{\scriptscriptstyle\rm BKT}
\right)^2,\hspace{1cm} T\rightarrow T_{\scriptscriptstyle\rm BKT}.
\end{equation}
Note also that this instability occurs when the fluctuations reach the
condition $\langle u^2\rangle\sim a_0\lambda$. This contrasts with the
often used Lindemann criterion for melting at $\langle
u^2\rangle=c_L^2 a_0^2$ and corresponds to a field dependent Lindemann
number $c_L\sim (B/B_\lambda)^{1/4}$ [see \cite{Ryu} where a
field-dependent $c_L$ was also found].

A precise treatment that can be used at all fields must include the
elastic distortions of the lattice within each layer. Within the
self-consistent harmonic approximation (SCHA) plus substrate model the
average energy cost for these distortions is given by a quadratic
form, integrated over all 2D modes in the Brillouin zone, $H^{\rm h}[{\bf
u}^0] =(1/8\pi^2)\int_{\scriptscriptstyle\rm BZ} d^2K\, u^0_i({\bf K})
\Phi^{ij}({\bf K}) u^0_j(-{\bf K})$ where
\begin{eqnarray}\label{eq:scha}
\Phi^{ij}({\bf K})\!&=&\!n_v\!\! \sum_{{\bf R}_\mu}\!\Biggl[\!
\left( 1\!-\!e^{i{\bf K}\cdot{\bf R}_\mu}\right)
\!\!\left\langle \partial_i\partial_jV_0
\right\rangle+ \sum_{n\ne 0}
\!\!\left\langle  \partial_i\partial_jV_n
\right\rangle\Biggr]\\
&&\hspace{-1cm}=\!
\varepsilon_0d
n_v^2\!\sum_{{\bf Q}_\mu}\!
\Biggl[
f_{ij}({\bf Q}_\mu\!+{\bf K})
\!-\!\!f_{ij}({\bf Q}_\mu)
\!+\!
\delta_{ij}
\frac{2\pi e^{-\frac{Q_\mu^2\langle u^2\rangle}2}}{1+\lambda^2Q_\mu^2}
\Biggr],\nonumber
\end{eqnarray}
with ${\bf R}_\mu$ and ${\bf Q}_\mu$ the real and reciprocal lattice
vectors and $f_{ij}({\bf Q})=Q_iQ_j(4\pi/Q^2) e^{-Q^2\langle
u^2\rangle/2}$.  The first two terms in (\ref{eq:scha}) are the
2D-elasticity and the last is the contribution from the substrate.
Again, we have ignored correlations in displacements $\langle
u^m_i({\bf R}_\mu) u^n_j({\bf R}_\nu) \rangle = \delta_{\mu\nu}
\delta_{mn}\delta_{ij}\langle u^2\rangle/2$ in the last line. Note
that the ${\bf Q}_\mu=0$ substrate term is cancelled by the ${\bf
Q}_\mu=0$ part of the second 2D-elastic term.  This is a reflection of
the long range divergences in the 2D system with log (Coulomb)
interactions, which do not exist for the 3D system of stacks where the
circulation currents are screened beyond $\lambda$.

The elastic matrix decomposes to transverse
\hbox{$(\delta_{ij}-\hat{K}_i\hat{K}_j) \Phi^{ij}({\bf K})=
c_{66}({\bf K})K^2+n_v\alpha_{\rm s}$} and longitudinal
\hbox{$\hat{K}_i\hat{K}_j \Phi^{ij}({\bf K})= c_{11}({\bf K})K^2+
n_v\alpha_{\rm s}$} projections, where $c_{66}$ and $c_{11}$ are the
dispersive shear and compression moduli, and $\alpha_{\rm s}$ is the
substrate strength,
\begin{equation}
  \label{eq:alpha_s}
\alpha_{\rm s}= 2\pi\varepsilon_0d n_v \sum_{{\bf Q}_\mu\ne 0} 
\frac{e^{-Q_\mu^2\langle u^2\rangle/2}} {1+\lambda^2Q_\mu^2}.
\end{equation}
The $\langle u^2\rangle\rightarrow 0$ limit of $c_{66}$ and $c_{11}$
recovers the usual form for the elasticity of a 2D vortex
lattice\cite{Brandt_Elastic}, with a shear modulus
$c^0_{66}=n_v\varepsilon_0d/4$ and a diverging compression modulus
$c^0_{11}(K)=4\pi\varepsilon_0dn_v^2/K^2$ at small $K$.  A finite
$\langle u^2\rangle$ softens these moduli, although the diverging
compression modulus remains.  To solve self-consistently for $\langle
u^2\rangle$ we use the equipartition result with the substrate,
\begin{eqnarray}
  \label{eq:2dequi}
  \langle u^2\rangle\!&=&\!\!\!
\int_{\scriptscriptstyle\rm BZ}\!
\frac{d^2 K}{(2\pi)^2}\!
\left[ \frac{T}{c_{66}({\bf K})K^2\!+\!n_v\alpha_{\rm s}}
+\frac{T}{c_{11}({\bf K})K^2\!+\!n_v\alpha_{\rm s}}
\right]\nonumber\\
&\approx& \frac{T}{4\pi c_{66}}\ln\left(1+
\frac{c_{66}K_{\scriptscriptstyle\rm BZ}^2}{n_v\alpha_{\rm s}}\right)
+\frac{T}{4\pi c_{11}(K_{\scriptscriptstyle\rm BZ}) +\alpha_{\rm s}},
\end{eqnarray}
where in the last line we have used the small $K$ limits for $c_{66}$
and $c_{11}$ and the circular-Brillouin-zone approximation with
$K_{\scriptscriptstyle\rm BZ}\approx\sqrt{4\pi}/a_0$.  Note how the
substrate potential cuts off the log divergence for the 2D shear
fluctuations.

\begin{figure}
\vspace{-0.3cm}
\centerline{\epsfxsize= 7.5cm\epsfbox{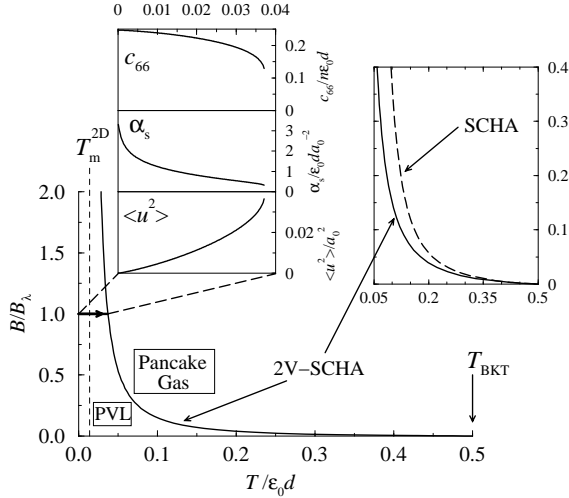}}
\caption{ The instability line for the PVL in the $B$-$T$ plane
calculated with the two-vertex-SCHA. The line goes asymptotically to
the 2D melting temperature at high fields, and ends at
$T_{\rm\scriptscriptstyle BKT}$ at zero field.  The left inset shows
the two-vertex-SCHA results for the shear modulus $c_{66}$, the
substrate strength $\alpha_{\rm s}$, and the pancake fluctuation width
$\langle u^2\rangle$ with increasing temperature at $B=B_\lambda$.
The right inset is a low-field zoom of the instability line showing
the result (dashed) of the simple SCHA scheme for comparison: the two
lines coincide at very low fields when the substrate dominates over
the 2D elasticity, but the SCHA overestimates the stability of the
lattice at higher fields.  We do not include here the intrinsic
temperature dependence in real superconductors of the penetration
depth and the energy scale $\varepsilon_0$ (but see Fig.~2).  }
\label{fig:tu}
\end{figure}

The self-consistent Eqs.\ (\ref{eq:scha}) and (\ref{eq:2dequi})
determine $\langle u^2\rangle$. Above a temperature $T_u$ there are no
solutions, giving an upper bound to the melting transition.  However,
in \cite{Dodgson_preprint} it was shown that the SCHA underestimates
the degree of anharmonic thermal softening of a 2D lattice because of
the neglect of odd terms in the anharmonicity. Such cubic
anharmonicities are included in the two-vertex-SCHA
\cite{Dodgson_preprint} giving results which compare well to numerical
simulations. In Fig.~\ref{fig:tu} we show the instability line in the
$B$-$T$ phase diagram using the two-vertex-SCHA. The line interpolates
between $T_{\scriptscriptstyle\rm BKT}$ at small fields and
$T^{\rm\scriptscriptstyle 2D}_m$ at high fields.  Also shown is the
significant thermal softening of both $c_{66}$ and $\alpha_{\rm s}$ at
$B=B_\lambda$ with the resulting anharmonic contributions (beyond
linear in $T$) for $\langle u^2\rangle$.

\vspace{-0.3cm} The above instability line only marks the upper
temperature limit of the solid phase. A true first-order transition
occurs when the free energies of two phases cross.  For melting
transitions it is often difficult to accurately determine the free
energy of the liquid. We can make progress for PVL melting, as the
liquid state behaves to a good approximation as uncoupled 2D liquids
in each layer. The free energy of a 2D liquid with log-interactions
can be extracted from numerical simulations\cite{Caillol}.  A crucial
ingredient is that the free energy is known exactly\cite{Alastuey} at
one special temperature $T=\varepsilon_0 d$. For the solid we
calculate the free energy of a 2D crystal on a self-consistent
substrate.  We must be careful to take the same normalization in both
phases: as in \cite{Alastuey} we normalize with respect to the ideal
gas of $N$ pancakes, $Z(H=0)=1$, and define the partition function,
\begin{equation}
Z=\frac1{N!}\int \prod_i \left(\frac {n_v}e\right) d^2 R_i\,\,
e^{-H[{\bf R}_i]/T},
\end{equation}
fixing the free energy, measured from the ideal gas, as $F=-T\ln
Z$. We write $H=H_{\rm\scriptscriptstyle 2D}+H_{\rm s}$ where
$H_{\rm\scriptscriptstyle 2D}= (N/2) \bigl[ \sum_{{\bf R}_\mu \ne 0}
V_0(R_\mu) -n_v \int d^2R V_0(R)\bigr]$ and $H_{\rm s}=(N/2) \bigl[
\sum_{{\bf R}_\mu, n\ne 0} V_n(R_\mu) +n_v \int d^2R V_0(R)\bigr]$ are
the 2D interaction term and substrate energy respectively\cite{Foot}.

In the solid, the free energy of harmonic fluctuations is
straightforward to calculate.  
The right-hand side of the 
inequality $F\le F^{\rm h} +\langle H- H^{\rm h}\rangle_{\rm h}$
is minimized by the SCHA (see \cite{Dodgson_preprint}) where $H^{\rm h}$ is
defined by (\ref{eq:scha}). The harmonic free energy of 2D fluctuations is
\begin{eqnarray}
F^{\rm h}&=&-\frac{NT}{2n_v} \int_{\rm BZ} \frac{d^2K}{(2\pi)^2}
\left[ \ln\left( \frac{\pi T n_v^2} {e(c_{66}K^2+n_v\alpha_{\rm s})}
\right)\right.\\
&&\left.\hspace{3cm}
+ \ln\left(\frac{\pi T n_v^2} {e(c_{11}K^2+n_v\alpha_{\rm s})}\right)
\right].\nonumber
\end{eqnarray}
The correction part of the variational free energy has contributions
from $H_{\rm\scriptscriptstyle 2D}$ and $H_{\rm s}$. Ignoring the small
anharmonic part of the 2D energy, the difference 
$\langle H_{\rm\scriptscriptstyle 2D}-
H_{\rm\scriptscriptstyle 2D}^{\rm h}\rangle_{\rm h}$
is just the ground state energy (which is not included in 
$H_{\rm\scriptscriptstyle 2D}^{\rm h}$).
In \cite{Alastuey} the energy
for a 2D lattice of log-interacting particles is found to be
$E_{\rm\scriptscriptstyle 2D}^0=0.749N\varepsilon_0d$. 
The substrate energy correction, $E_{\rm s}=\langle
H_{\rm s} - H_{\rm s}^{\rm h}\rangle_{\rm h}$,
is
\begin{eqnarray}
E_{\rm s} &=&-N\!\!\left[ 2\pi n_v \varepsilon_0d\!\!\sum_{{\bf Q}_\mu\ne 0}
\frac{e^{-Q_\mu^2 \langle u^2\rangle/2}}{Q_\mu^2(1+\lambda^2Q_\mu^2)}
\!+\!\frac12 \alpha_s \langle u^2\rangle \right]\!,
\end{eqnarray}
and the sum $F^{\rm h}+E_{\rm\scriptscriptstyle 2D}^0+E_{\rm s}$
is the variational free energy.

\begin{figure}
\vspace{-0.2cm}
\centerline{\epsfxsize= 8cm\epsfbox{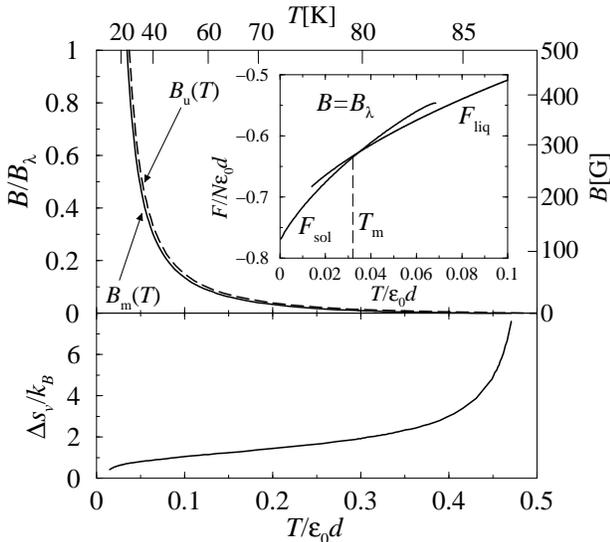}}
\vspace{0.1cm}
\caption{The full line in the upper graph shows the melting line
$B_m(T)$ of the pancake vortex lattice as calculated by comparing the
free energies of the solid and liquid phases. The dashed line is the
stability limit $B_u(T)$ of the PVL as shown in Fig.\ \ref{fig:tu}.
The inset shows the free energy comparison at $B=B_\lambda$. The lower
graph gives the entropy jump per pancake vortex $\Delta s_v$.  Also
shown are the real scales of $T$ and $B$ assuming
$\lambda(T)=\lambda(0)/[1-(T/T_{c0})^2]^\frac12$, with
$\lambda(0)\approx 2000$\AA, $d\approx 15${\AA} and $T_{c0}\approx
100$~K.  The low-field reentrant melting line, not shown here, will
cut $B_m(T)$ below $10^{-2}B_\lambda$.  }
\label{fig:comp_lines}
\end{figure}

In the liquid phase, $\langle H_{\rm s}\rangle$ is very small and we
should find the free energy of a 2D liquid. The internal energy
$U(\Gamma)$ (with $\Gamma=2\varepsilon_0 d/T$) is found from
simulations \cite{Caillol},
\begin{equation}\label{eq:ugamma}
U(\Gamma)\!=\!\left(-0.751+ 0.880\, \Gamma^{-0.74}\!\!-0.209\,
\Gamma^{-1.7} \right) N\varepsilon_0d.
\end{equation}
From the relation $F=U+T\partial_TF$, the excess free energy can be
written as $\Gamma F_{\rm liq}(\Gamma)=\Gamma_0 F_{\rm liq}(\Gamma_0)
+\int^\Gamma_{\Gamma_0} U(\Gamma')d\Gamma'$.  We know the value at
$\Gamma_0=2$, where $F_{\rm liq}(2)= 0.081\varepsilon_0d N$
\cite{Alastuey}, so we can integrate (\ref{eq:ugamma}) to give (with
$t=T/\varepsilon_0 d$),
\begin{equation}\label{eq:fliquid}
F_{\rm liq}\!=\!\left(-0.751 \!-\!\!1.287 t \!+\!2.027 t^{0.74}\!\!
+\!0.092 t^{1.7}\right)\!  N \varepsilon_0d.
\end{equation}
Using these results we can plot the free energy of both phases and see
where they cross; a typical example at $B=B_\lambda$ is shown in the
inset of Fig.~\ref{fig:comp_lines}. Calculating the crossing point
numerically at different fields, we find a melting line (see
Fig.~\ref{fig:comp_lines}) just below the stability limit of the
lattice that approaches zero field at $T\approx 0.47 \varepsilon_0d$.

The jump in slope $S=-\partial_T F$ gives a latent heat $T\Delta
S=\Delta U$. In Fig.\ \ref{fig:comp_lines} we plot the entropy jump
per pancake vortex as a function of the transition temperature. At
high fields (small $T$) it approaches the value $\Delta s_v\approx 0.4
k_{\rm\scriptscriptstyle B}$, consistent with 2D simulations
\cite{Caillol} and simple estimates\cite{Dodgson_Jumps}. At low fields
($T\sim0.5\varepsilon_0d$) the latent heat appears to weakly diverge
as $T\rightarrow T_{\rm\scriptscriptstyle BKT}$. We understand this as
the energy of the liquid is roughly constant, $U(T\approx
T_{\rm\scriptscriptstyle BKT})\approx -0.219 N\varepsilon_0 d$, while
the energy of the solid is dominated by the substrate term, $E_{\rm
s}\sim N\varepsilon_0d \ln(\lambda/a_0)$ so that the latent heat is
$\Delta U\sim -E_{\rm s}$. This gives an entropy jump per vortex pancake
\hbox{$\Delta s_v=\Delta U/NT \sim k_{\rm\scriptscriptstyle B}
\ln(B_\lambda/B)$}.  This is of the same form as the entropy
difference between an ideal gas, $\Delta s_v^{\rm gas}\approx
k_{\rm\scriptscriptstyle B} \ln(a_0^2/\xi^2)$ and the (reduced
phase-space) solid $\Delta s_v^{\rm sol}\approx
k_{\rm\scriptscriptstyle B} \ln(\langle u^2\rangle/\xi^2)$ when
$\langle u^2\rangle \sim a_0\lambda$ (as found above for the low-field
instability).  We do not include here the $T$ dependence of $\lambda$
in real systems that gives extra terms in the latent heat\cite{Dodgson_Jumps}.

\vspace{-0.2cm} Previously, melting of the magnetically coupled PVL
has been analyzed numerically\cite{Reefman} and via density functional
theory (DFT) of the liquid phase\cite{DFT}. The early simulations in
\cite{Reefman} find evidence of melting at $T\approx 0.09\varepsilon_0
d$ when $B/B_\lambda\approx 0.2$ (close to our melting line). The DFT
gives the stability limit of the liquid and provides results
consistent with ours at fields above $0.5 B_\lambda$. At lower fields
the DFT gives a liquid instability above our melting line, a
discrepancy which requires further study.

To compare to real superconductors our units of field $B_\lambda$ and
of temperature $\varepsilon_0 d$ must be scaled due to the intrinsic
variation in $\lambda(T)$ (diverging at a temperature $T_c^0$). Doing
this, we find that our melting curve of Fig.~\ref{fig:comp_lines} lies
below experimental melting lines \cite{Zeldov,Lee} for reasonable
choices of $\lambda(T)$; this is because the Josephson coupling,
neglected in this paper, becomes important as $T_c^0$ is
approached\cite{BlatterGLN} and stiffens the vortex lattice. Our
results may be recovered in experiments by suppressing the Josephson
coupling with a strong in-plane magnetic field.

We thank Misha Feigelman and Lev Bulaevskii for useful discussions.
Work in Z\"urich was financially supported by the Swiss National
Foundation and in Argonne by the NSF Office of the Science and
Technology Center under contract DMR-91-20000 and the US.\ DOE,
BES-Materials Sciences, under contract W-31-109-ENG-38.

\end{multicols}


\begin{thebibliography}{99}
\bibitem{Foot} 
Note that we add and subtract the background energy from a continuous 
opposite charge, as the diverging
energy of the 2D log-system must cancel for 3D stacks.
\end{thebibliography}
\end{document}